\documentclass[12pt]{iopart}

\usepackage{iopams}
\usepackage{graphicx}
\usepackage{stmaryrd}
\expandafter\let\csname equation*\endcsname\relax
\expandafter\let\csname endequation*\endcsname\relax
\usepackage{mathtools}
\usepackage[dvipsnames]{xcolor}
\renewcommand{\vec}[1]{\boldsymbol{#1}}

\newcommand{\caA}{{\mathcal A}}
\newcommand{\caB}{{\mathcal B}}

\newcommand{\caD}{{\mathcal D}}

\newcommand{\caK}{{\mathcal K}}

\newcommand{\caO}{{\mathcal O}}

\newcommand{\caQ}{{\mathcal Q}}

\newcommand{\caS}{{\mathcal S}}

\newcommand{\caU}{{\mathcal U}}

\newcommand{\diff}{d} 

\newcommand{\mean}[1]{{\left< #1 \right>}}
\newcommand{\genL}{{\mathbb L}}

\begin{document}

\title{Nonequilibrium temperature response for stochastic overdamped systems}

\author{G Falasco$^{1}$ and M Baiesi$^{2,3}$}

\address{$^1$Max Planck Institute for Mathematics in the Sciences, Inselstr. 22,
04103 Leipzig, Germany}
\ead{gianmaria.falasco@mis.mpg.de}

\address{$^2$ Dipartimento di Fisica ed Astronomia, Universit\`a di Padova, Via Marzolo 8, I-35131 Padova, Italy}
\address{$^3$ INFN, Sezione di Padova, Via Marzolo 8, I-35131 Padova, Italy}
\ead{baiesi@pd.infn.it}

\begin{abstract}
  The thermal response of nonequilibrium systems requires
  the knowledge of concepts that go beyond entropy production.
  This is showed for systems obeying overdamped Langevin dynamics, 
  either in steady states or going through a relaxation process.
  Namely, we derive the linear response to perturbations of the noise
  intensity, mapping it onto the quadratic response to a constant
  small force. The latter, displaying divergent terms, is explicitly
  regularized with a novel path-integral method.
  The nonequilibrium equivalents of heat capacity and thermal
  expansion coefficient are two applications of this approach,
  as we show with numerical examples.
\end{abstract}


\section{Introduction}

The determination of response functions is arguably one of the most topical issues in statistical physics.
Even though its history dates back to the works of Einstein, Nyquist and Onsager~\cite{ein05, nyq28, ons31,ons31b}, it was Kubo \cite{kub57,tod92} who subsumed the later developments \cite{cal51, gre52, gre54} under a general theory. For a system slightly driven off equilibrium, the Kubo formula gives the linear response of an observable in terms of the equilibrium time-correlation between the observable itself and the entropy produced by the perturbation. 
The first systematic application of Kubo's theory ---along with kinetic theories based on generalized Boltzmann equations--- underscored the endeavor to calculate the transport coefficients of moderately dense gasses \cite{ern69}. These efforts culminated in
the discovery of the algebraic decay in time of the correlation functions entering Kubo formulas~\cite{ald70,dor70,zwa70}, which prevents the existence of transport coefficients in low dimensions.

Later, the possibility to perform progressively more efficient computer simulations and thus to compute  response functions numerically, led to the extension of the original theory to thermostatted systems arbitrarily perturbed from an initial equilibrium state \cite{eva90}. Remarkably, it was established that the (nonlinear) response to an external driving is largely insensitive to the choice of the thermostatting mechanisms \cite{eva93b}, represented by the artificial forces required to maintain nonequilibrium steady-state conditions \cite{ron07}. 

In contrast to such major achievements, the related theory for the response upon perturbation of nonequilibrium states has progressed far more slowly. Apart from the obvious obstacle represented by the lack of knowledge of nonequilibrium phase-space distributions, further difficulties are met when dealing rigorously with deterministic dynamical systems, owing to the fractal nature of their invariant distribution \cite{rue09, mar08,col12,col14}. 
Nonequilibrium response theories have rather flourished for stochastic dynamics \cite{han78, fal90,cug94,rue98,nak08,che08,spe06,spe09,sei10,pro09,ver11,lip05,lip07,bai09,bai09b}, which is applicable to a wide variety of complex systems in physics as well as in related sciences. 
However, most of these results are usually restricted to {\em mechanical} perturbations and do not consider {\em thermal} perturbations. Thus, they do not allow one to compute quantities such as nonequilibrium heat capacities and thermal expansions coefficients, which would arise as the (integrated) linear response to step variations of the temperature, i.e., of the noise intensity in the stochastic dynamical equations.  Besides some previous formal results \cite{che09b,sei10}, only recently there appeared formulas for the thermal response of driven stochastic systems, which are given in terms of correlations between state observables  calculated in the unperturbed state. Apparently, the mathematical difficulties entailed by handling noise variations  require either to introduce an explicit time-discretization to avoid divergences in the response \cite{bai14,yol15} or to rely on a rescaling of the stochastic dynamics in order to derive regular results \cite{fal15b}. 

The present work is devoted to show that neither of these expedients is actually necessary. A well-defined thermal response formula can be derived by standard path integral techniques, in close analogy to the case of deterministic perturbations.
After introducing the model equations in section~\ref{sec:model}, we define in section~\ref{sec:path}  the  linear response to a temperature perturbation of a generic observable of the system. In section~\ref{sec:map} after a brief explanation of the formal differences from the ordinary response to a deterministic forcing, we tackle the problem first showing that the thermal response is equivalent to a portion of the quadratic (i.e.~second-order) response to a constant force. Such expression, which displays divergent terms, is then explicitly regularized in section~\ref{sec:reg} and is showed to be equivalent to a Kubo formula in equilibrium. In section~\ref{sec:num} we illustrate two applications of these results: the energy susceptibility of a driven quenched particle (that is the non-equilibrium specific heat for zero driving)  and the thermal expansion coefficient of an anharmonic lattice subjected to large heat flows. Moreover, in the simplest tractable case of a freely diffusing particle we connect our formulas to the Einstein relation. A summary and an outlook are finally given in the conclusions.

\section{Overdamped Langevin dynamics}
\label{sec:model}
The overdamped diffusive system we consider consists of $N$ degrees of freedom, denoted $\vec x = \{x_1,\dots x_N \}$. For instance, $x_j$ may be a component of a particle position vector in $d$-dimensions, so that $N= n d$ if the system is composed by $n$ particles. The dynamics is given by the overdamped Langevin equation
\begin{align}\label{ods}
\dot x_j(t) = \mu_j F_j(\vec x(t)) + \sqrt{2 \mu_j T_j} \xi_j(t),
\end{align}
where each Gaussian white noise $\xi_j$ is uncorrelated from the others,
\begin{align}\label{noise}
\mean{\xi_j(t)\xi_{j'}(t')} =  \delta(t-t') \delta_{jj'} .
\end{align}
The $j$-th bath temperature $T_j$ and mobility $\mu_j$ (which is the inverse of a damping constant) determine the strength of the noise term, while the drift depends on $\mu_j$ and on the mechanical force $F_j(\vec x(t))$. Such structure respects local detailed balance and thus assumes that the baths are noninteracting with each other and always in equilibrium, regardless of the nonequilibrium conditions experienced by the system. Temperatures and mobilities in our formalism do not depend on the coordinates, hence there is no ambiguity in the interpretation of the 
stochastic equation. Throughout this paper we will always consider the Stratonovich convention, that is the midpoint rule is employed to discretise in time \eqref{ods}~\cite{sek10}. Hence none of the integrals will be of the Ito type and the rules of standard calculus can be applied.

In this context of temperature response, even if equations have a noise prefactor that does not depend on the system's state $\vec x$, it turns out that the choice of using Stratonovich path-weights rather than Ito ones is not trivial.
As discussed previously~\cite{yol15}, by differentiating with respect to temperature one proves a response formula that depends on the choice of the path-weight. One can check that the formulas in this paper are indeed different from those found adopting the Ito convention~\cite{bai14}. Ultimately, the path-weight, and thus the corresponding discretization of \eqref{ods}, have to be chosen consistently with the physical process that \eqref{ods} is meant to model. The Ito convention, for example, is by construction suitable for numerical data generated by integration of \eqref{ods} with the Euler scheme~\cite{bai14}. On the other hand, the Wong-Zakai theorem \cite{oksendal} ensures the Stratonovich convention to be adequate to experimental data, for which the white noise is an idealized limit of the short correlation times of the microscopic degrees of freedom.

The $F_i$'s are generic nonconservative forces that may bring the system arbitrarily far from equilibrium.
In the resulting statistical averages, denoted $\mean{\ldots}$, there is an understood dependence
on the initial density of states $\rho_0(\vec x_0)$, with $\vec x_0= \vec x(0)$. This may coincide or not with the steady state density.
Finally, we introduce the backward generator of the Markovian dynamics \eqref{ods}, written as a sum of ``one-coordinate" operators~$\genL_j$,
\begin{align}\label{L}
\genL = \sum_{j=1}^N \genL_j
\quad \textrm{with} \quad
\genL_j = \mu_j F_j(\vec x) \partial_{j} + \mu_jT_j \partial^2_{j},
\end{align}
where  we set $\partial_{x_j} \equiv \partial_j$ to avoid clutter.
It gives the average time derivative of a state observable $\caO(t)$ as $\frac{\diff }{\diff t} \mean{\caO(t)}= \mean{\genL \caO(t)}$. Hereafter for any state observable we use the shorthand notation $\caO(\vec x(t),t) \equiv \caO(t)$ to indicate the implicit (and possibly explicit) dependence on the time $t$.

\section{Linear response in path integral formalism}
\label{sec:path}

We imagine to perturb the system \eqref{ods} varying the noise amplitude through a time dependent parameter ${\theta(t) \ll 1}$ switched on at time $t=0$, namely 
\begin{align}\label{Theta}
T_i\, \to\, \Theta_i(t)\equiv T_i + \epsilon_i \theta(t),
\end{align}
where $\epsilon_i$ is a constant determining the $i$-th amplitude of the perturbation. 
This renders \eqref{ods} for a perturbed degree of freedom into the form 
\begin{align}\label{ods_h}
\dot x_i(t) = \mu_i F_i(\vec x(t)) + \sqrt{2 \mu_i \Theta_i(t)} \xi_i(t).
\end{align}
Without loss of generality we assume the mobility to be independent of temperature. The extension to the case where $\mu_i=\mu_i(\Theta_i)$ does not involve particular difficulties, since the linear response would be just the sum of the temperature response here described plus a standard response to a deterministic perturbation~\cite{bai09,bai09b}, which arises linearising the term~$\mu_i F_i$.

The aim is to calculate the linear response of a generic observable $\caO(t)$ to the just introduced temperature change, defined by
\begin{align}\label{resp}
R_{\caO,\theta}(t,t')
\equiv \frac{\delta\mean{ \caO(t)}_{\theta}}{\delta \theta(t')}\Bigg|_{\theta=0}
= \frac{\delta}{\delta \theta(t')} \int \caD \vec x_{\theta} \diff \vec x_0  \caO(t) P_{\theta}[\vec x] \rho_0(\vec x_0)\Bigg|_{\theta=0} .
\end{align}

Here $\mean{\dots}_{\theta}$ denotes an average performed in the perturbed dynamics \eqref{ods_h} starting from the state $\rho_0(\vec x_0)$, which is unaltered by the perturbation. The associated path weight, proportional to the probability of a trajectory $[\vec x] \equiv \{\vec x(s): 0 \leqslant s \leqslant t\}$ solution of \eqref{ods_h}, is expressed as \cite{zinn02}
\begin{align}\label{path}
P_{\theta}[\vec x]= \exp \caA_{\theta}[\vec x],
\end{align}
with the action functional
\begin{align}\label{action}
\caA_{\theta}[\vec x] = 
-\sum_{j=1}^{N} \bigg\{\int_0^t \diff s   \frac{(\dot x_j(s) - \mu_j F_j(s))^2}{4\mu_j \Theta_j(s)}
 + \frac {\mu_j} 2 \int_0^t \diff s \partial_j F_j(s) \bigg\}.
\end{align}
The last term in \eqref{action} appears as the functional Jacobian in deriving the the path-weight for $[\vec x]$ from the Gaussian path-weight associated to the noise $\xi_i$, and depends on the convention used to discretise \eqref{ods_h} (e.g. it would be absent with the Ito convention). In the following we will also make use of the unperturbed action $\caA\equiv \caA_{\theta}|_{\theta=0}$, which amounts to replace $\Theta_j$ with $T_j$ in \eqref{action}. 

Deep physical insights come from separating any action of the form \eqref{action} 
into time-antisymmetric ($\caS$) and time-symmetric ($\caK, \caK_0$) components:
\begin{align}\label{Asmart}
\caA[\vec x] &= \frac 1 2 \caS[\vec x] -  \caK[\vec x] -  \caK_0[\vec x]
\end{align}
with
\begin{align}
\label{term-act-S}
\caS[\vec x]   & \equiv \sum_{j=1}^N \frac 1 {T_j}\int_0^t \diff s F_j(s) \dot x_j(s)\,,\\
\label{term-act-K}
\caK[\vec x]   & \equiv \sum_{j=1}^N \int_0^t \diff s 
  \frac{\mu_j}{4 T_j} \left[ F_j^2(s)+ 2 T_j \partial_j F_j(s) \right],\\
\label{term-act-K'}
\caK_0[\vec x]  & \equiv \sum_{j=1}^N \int_0^t \diff s   \frac{\dot x_j^2(s)}{4\mu_j T_j} \,.
\end{align}
The integrated entropy flux $\caS[\vec x]$ is the antisymmetric part of the action $\caA$ under the time-reversal transformation $x_j(s) \to x_j(t-s)$. It is defined consistently with thermodynamics as the sum of the individual heat fluxes into the reservoirs, each weighted by the respective bath temperature~\cite{sek10}.
The time-symmetric terms have been studied in connection with the notion of dynamical activity, formerly introduced in the context of jump systems~\cite{lec05,mer05,gar09}, where it counts the number of jumps and provides important informations, e.g., on the state of glassy systems. Both $\caK[\vec x]$ and $\caK_0[\vec x]$ in fact may quantify an amount of activity in the diffusive system we are considering~\cite{ful13}. Being $\caK_0[\vec x]$ related to the mean square displacement of the $N$ degrees of freedom, it offers a direct estimate of the trajectory frenzy. Nevertheless, this kinetic-like term should be understood as part of the functional measure \cite[Sec.~2.2]{zinn02}, as it selects from all possible trajectories the Brownian paths that make $\caK_0$ finite in the limit $ds \to 0$ (i.e. those that satisfy $d x_j^2 \sim ds$). The functionals $\caS$ and $\caK$ are then the statistical weights of such selected trajectories.
Therefore, in the following we will reserve the name {\em dynamical activity} for $\caK$, which was shown to be a good measure of the system activity~\cite{ful13}. Written as
\begin{align}
\caK[\vec x] &\equiv \int_0^t \diff s V_{\rm eff}(\vec x(s)),
\end{align}
it may be seen as a time-integral of a state variable $V_{\rm eff}(\vec x)$ that, for systems with interactions deriving from  an energy potential $U(\vec x)$ and with a global bath temperature $T$, would read 
\begin{equation}
V_{\rm eff}(\vec x) = \frac{1}{4 T} \sum_j \mu_j \left[ (\partial_jU(\vec x))^2- 2 T \partial^2_j U(\vec x) \right].
\end{equation}
Such quantity was called effective potential~\cite{aut09,pit11} and is proportional to the escape rate from a configuration $\vec x$, as the probability to remain in $\vec x$ for a short time $\Delta t$ is $\sim \exp( -V_{\rm eff} \Delta t)$.
For our nonequilibrium systems we generalise such concept by writing $V_{\rm eff}(s) = \sum_{j=1}^N \lambda_j(s) $,
with
\begin{align}
\lambda_j(s)\equiv  \frac{\mu_j}{4 T_j} \left[ F_j^2(s)+ 2 T_j \partial_j F_j(s) \right].
\end{align}
The escape rate of the degree of freedom $x_j$, denoted $\lambda_j$, follows from evaluating the action at fixed $\vec x$ along a very short trajectory of duration $\Delta t \ll 1$, that is,
$\lim_{\Delta t\to 0}\text{Prob}(\vec x,s+ \Delta t| \vec x, s)/\Delta t =\exp(-\sum_{j=1}^N \lambda_j(s))$.

In the following sections we will sometimes also use the name {\em frenesy} for describing correlation functions in the response formulas involving time-symmetrical features. This alternate naming originated in the response-theory framework~\cite{bai13} and usually refers to quantities akin to $\caK$ ---more specifically, to its excess generated by a perturbing force---, namely to quantities assessing the system {\em impatience} for changing its state (rather than direct measures of the trajectory zigzags). Hopefully the double terminology is guiding the reader through the connections with the recent literature.

\section{Response to heating as response to a force}
\label{sec:map}

We are now in the position to develop the thermal linear response theory, but we immediately find an obstacle.
Since the path weight \eqref{path} is normalised to one, $\int \caD \vec x_{\theta} P_{\theta}[\vec x]=1$, the functional measure $\caD \vec x_{\theta}$ in \eqref{resp} contains the noise temperatures $\Theta_j$ (see e.g.~\cite{zinn02, lau07}), and therefore depends itself on the external parameter $\theta$. This is a major difference with respect to an external perturbation of the deterministic forces, which leads to the formal difficulties reported in \cite{bai14}, namely the introduction of an explicit time-mesh to avoid singularities in the results.
To overcome this problem we first seek a more manageable expression for the path average. That is obtained through an Hubbard-Stratonovich transformation \cite{negele88} of the action that, introducing an auxiliary variable $\vec y$, linearises the quadratic term in \eqref{action} and removes the $\theta$ dependence from the functional measure of the path weight (see e.g. \cite{lau07}). By doing so, it is easy to bring~\eqref{resp} in the form (see~\ref{sec:HS}) 
\begin{align}
 &\frac{\delta\mean{ \caO(t)}_{\theta}}{\delta \theta(t')}\Bigg|_{\theta=0} 
 = \sum_{i}\frac{\epsilon_i}{\mu_i}R^{(2)}_{\caO,f_i}(t,t',t'). \label{R2}
\end{align}
Here $R_{\caO,f_i}^{(2)}$ is the second-order response function to a constant force perturbation $f_i$ of the $i$-th degree of freedom~\cite{lip08}, namely  
\begin{align}
R^{(2)}_{\caO,f_i}(t,t',t'') \equiv \frac{\delta^2 \mean{\caO(t)}_{\vec f}}{\delta f_i(t')\delta f_i(t'')}\Bigg|_{\vec f=0},
\end{align}
where $\mean{\dots}_{\vec f}$ now denotes the average with respect to the perturbed dynamics 
\begin{align}\label{const}
\dot x_i = \mu_i (F_i(\vec x)+f_i )+ \sqrt{2 \mu_i T_i} \xi_i.
\end{align}
Formal calculation of response functions to external forces poses no technical difficulty ~\cite{cug94,lip05,bai09}. After integrating out the auxiliary variable $\vec y$, it is straightforward to find for \eqref{R2}
\begin{align}
R^{(2)}_{\caO,f_i}(t,t',t')
&= \frac{1}{2 T_i}  \frac{\delta}{\delta  f_i(t')} \mean{(\dot x_i(t')- \mu_i F_i(t') -\mu_i f_i(t')) \caO(t)}_{\vec f} \bigg|_{\vec f=0} \nonumber \\
&=  \frac{1}{4 T_i^2} \Big[ \mean{(\dot x_i(t') - \mu_i F_i(t'))^2\caO(t) } - 2\mu_i T_i \delta(0) \mean{\caO(t)}  \Big]. \label{delta}
\end{align}
Summing up, a standard Hubbard-Stratonovich transformation has allowed us to write the linear response of an observable $\caO$ to a temperature change as the second-order response to a state-independent force, thus arriving at the intermediate result
\begin{align}\label{response}
R_{\caO,\theta}(t,t')= \sum_i \frac{\epsilon_i}{4 \mu_i T_i^2} \Big[ & 
\Big<\caO(t) \big(\dot x_i^2(t') -2\mu_i  \dot x_i(t') F_i(t') 
 + \mu_i^2 F_i^2(t')\big) \Big> - 2 \mu_i T_i \delta(0)\mean{\caO(t)} \Big].  
\end{align}
As anticipated, this result is slightly different from that of a previous 
approach~\cite{bai14} where the Ito convention was adopted
for the path-integrals.

\section{Regularization of the response}
\label{sec:reg}
In~\eqref{response} the divergence caused by the Dirac delta formally compensates the divergence in the squared velocity. This can be heuristically understood recalling that~\eqref{response}, despite being formally expressed in continuous time notation, can be interpreted in terms of discrete, albeit small, time intervals $\Delta t$ \cite{zinn02, gro98}. Therefore one has $\dot x_i^2\sim 1/\Delta t$, being the dynamics diffusive at short times, and clearly $\delta(0) \sim 1/\Delta t$. However, it would be convenient to recast~\eqref{response} as an explicit result devoid of singular terms. In the following we perform such operation, first for a single degree of freedom ($N=1$), and then extending the result to arbitrary $N$.

\subsection{One degree of freedom}
\label{sec:one-dof}
With one degree of freedom the parameter $\epsilon_i$ is superfluous and is thus set to $1$.
We first focus on the kinetic-like term by starting with the rewriting (valid for $t > t'$)
\begin{align}\label{diff}
\mean{\dot x^2(t') \caO(t)}=\frac{1}{2}\frac{\diff^2 \phantom{c}}{\diff t'^2}\mean{x^2(t') \caO(t)} -\mean{\ddot x(t') x(t') \caO(t)},
\end{align}
and by seeking a replacement for the correlation function $\mean{\ddot x(t') x(t') \caO(t)}$. This can be achieved recalling that the integral of a total derivative involving the path weight is null. Therefore, we may exploit the identity
\begin{align}\label{identity}
0 = \int \caD x \frac{\delta}{\delta x(t')} \caB[x] P[x]  =\mean{ \frac{\delta \caB}{\delta x(t')}} + \mean{ \caB \frac{\delta \caA}{\delta x(t')} }\,,
\end{align}
where $\caB$ is any functional of $\{x(s):0 \leqslant s \leqslant t \}$, and $\caA[x]$ is the unperturbed action 
\begin{equation}\label{S}
\caA[x]=-\frac{1}{4\mu T}\int_0^t \diff s (\dot x(s) - \mu F(s))^2 - \frac \mu 2 \int_0^t \diff s \partial_x F(s)\,,
\end{equation}
corresponding to \eqref{action} calculated at $\theta=0$, with $N=1$.
First, we evaluate the second term in~\eqref{identity} making use of the expression for the functional variation of the action derived in~\ref{sec:path_rel}, see~\eqref{derivative}. The entropy variation is shown to vanish, while the variation of $\caK[\vec x]$ expressed in terms of the backward generator $\genL$ gives 
\begin{align}
 \mean{ \caB \frac{\delta \caA}{\delta x(t')} }= 
 \mean{ \caB \frac{\delta \caK}{\delta x(t')} }=  
\frac{1}{2\mu T}  \mean{ \caB \Big[ \ddot x(t')- \mu \genL F(t')\Big]  }\label{deS}\,.
\end{align}
Hereafter we restrict to the case in which $F$ does not depend explicitly on time, but only via $x$. 
In order to extract from~\eqref{deS} the sought substitute for $\mean{\ddot x(t') x(t') \caO(t)}$,  
we choose $\caB= \caO(t) x(t')$ and the first term in~\eqref{identity} becomes
\begin{align}\label{diffA}
 \mean{ \frac{\delta \caB}{\delta x(t')}}= \mean{\frac{\delta \caO(t)}{ \delta x(t')} x(t')} + \mean{\caO(t)}\delta(0).
\end{align}
If $\caO$ is a state observable, i.e., it depends only on the trajectory endpoint, the first term on the right hand side of~\eqref{diffA} drops for all $t' \neq t$, since it reads $\frac{\delta \caO(t)}{ \delta x(t')}= \partial_x \caO(t) \delta(t-t')$.
Putting all the pieces together
we get the compact expression
\begin{align}
\mean{\ddot x(t') x(t') \caO(t)}& =  \mu \mean{\caO(t) x(t') \genL F(t')} -2\mu T \delta(0),
\end{align} 
which, plugged in the response formula~\eqref{response}, gives finally
\begin{align}
\label{R1}
R_{\caO,\theta}(t,t')  = \frac{1}{4 T^2}\bigg[  
& \frac{1}{2 \mu}\frac{\diff^2 \phantom{c}}{\diff t'^2}\mean{\caO(t) x^2(t')}+  \mean{\caO(t) \mu F^2(t')} \nonumber\\
& -\mean{\caO(t)  x(t')\genL F(t') } 
  -2\mean{\caO(t)  \dot x(t') F(t')} 
\bigg].
\end{align}
for $t' < t$.
This is a regularised  version of \eqref{response} valid for $N=1$ and any state observable $\caO$. We have traded the kinetic-like term and the Dirac delta in  \eqref{response} with a second-order time derivative and a correlation involving the backward generator. The second-order time derivative, even tough unusual for a linear response formula (but not for a second-order response function \cite{lip08}), is indeed necessary to obtain the correct result, as it can be easily verified in the analytically solvable case of a particle in free diffusion (see section~\ref{sec:free}). 

If one is interested in the response of path-dependent observables (namely, $\caO$ is a functional of the trajectory up to time $t$), the first summand in \eqref{diffA} is non-zero and hence~\eqref{R1} has to be supplemented by the term $- 2 \mu T \mean{\frac{\delta \caO(t)}{ \delta x(t')} x(t')} $. As an example we may consider the heat exchanged with the thermal bath in a time $t$, $\caQ[x]\equiv\int_0^t \diff s F(s) \dot x(s)$. It turns out that the response formula \eqref{R1} requires no additional term in this case, since
\begin{align}
\frac{\delta \caQ(t)}{ \delta x(t')}&=\partial_x F(t') \dot x(t')+ \int_0^t \diff s \dot \delta(s-t') F(s) \nonumber\\
&=\partial_x F(t') \dot x(t')- \partial_x F(t') \dot x(t')=0\,. \label{Q1}
\end{align}

\subsection{Many degrees of freedom}
\label{sec:many-dof}
The procedure  is easily extended to a system composed of $N>1$ degrees of freedom. Equations~\eqref{diff},~ \eqref{identity} and \eqref{diffA} are still valid replacing $x$ with $x_i$, and taking the action (corresponding to \eqref{action} calculated at $\theta=0$) 
\begin{align}\label{action_un}
\caA[\vec x] = &
-\sum_{j=1}^{N} \bigg\{ \frac{1}{4\mu_j T_j}\int_0^t \diff s (\dot x_j(s) - \mu_j F_j(s))^2
 + \frac {\mu_j} 2 \int_0^t \diff s \partial_j F_j(s) \bigg\},
\end{align}
where we reverted to the notation accommodating the particle labels. Equation~\eqref{deS} is then generalised to (see~\ref{sec:path_rel})
\begin{align}
\mean{ \caB \frac{\delta \caA}{\delta x_i(t')} }  &= \frac{1}{2 \mu_i T_i}  \mean{ \caB \ddot x_i(t')} 
- \mean{ \caB \frac{\delta \caK }{\delta x_i(t')}  }
 + \frac 1 2 \mean{ \caB \frac{\delta \caS }{\delta x_i(t')}  }.
\label{noifU}
\end{align}
In the following we focus on systems with two-body potential interactions, deferring the more general result (valid for arbitrary $d$, generic driving and interactions) to~\ref{sec:path_rel}. Yet, the results reported here are general enough to describe the thermal response of a broad class of non-equilibrium systems, such as heat conducting lattices in contact with different heat baths [Eq.~\eqref{many_dof}], and aging systems [Eq.~\eqref{R3}]. Under the above assumption,  the variation of $\caK[\vec x]$ in~\eqref{noifU}  is given by
\begin{align}\label{varK}
 \frac{\delta \caK }{\delta x_i(t')} = \genL^{(T_i)}F_i(t'),
\end{align}
where we identified the operator 
\begin{align}
\label{genLT}
\genL^{(T_i)} & \equiv \sum_{j=1}^N \frac{T_i}{T_j} \genL_j
\, =  \sum_{j=1}^N \left( \frac{T_i}{T_j} \mu_j \partial_j + T_i  \partial_j^2 \right)
\end{align}
which acts on the observables as if all temperatures were equal to $T_i$ and all forces $F_j$ were rescaled by $T_i/T_j$. Interesting, this rescaling is found by rewriting the Langevin dynamics in terms of a new time variable, the \emph{thermal time} $\tau_j \equiv t \frac{T_j}{T_i}$, by which \eqref{ods} reads
\begin{align}\label{odstau}
\frac{\diff x_j}{\diff \tau_j} = \mu_j \frac{T_i}{T_j} F_j + \sqrt{2 \mu_j T_i} \xi_j.
\end{align}
While $\genL$ is the generator of the stochastic dynamics in the kinematic time $t$, in view of \eqref{odstau}, the operator $\genL^{(T_i)}$ acts as the generator of the corresponding dynamics in thermal time coordinates. This permits to rationalize the variation of the dynamical activity~\eqref{varK} as the tendency to change $F_i$ measured with respect to the thermal time.

Coming back to the regularization of \eqref{response} we operate as before. We choose $\caB= \caO(t) x_i(t')$ and obtain, by means of~\eqref{identity},~\eqref{diffA} and \eqref{noifU}
\begin{align}
\mean{\caO(t)\ddot x_i(t') x_i(t')}= & \mu_i  \mean{ \caO(t) x_i(t') \genL^{(T)} F_i(t') } 
+  \mu_i T_i \mean{ \caO (t) x_i(t')  \frac{\delta \caS }{\delta x_i(t')}  }  -2\mu_i T_i \delta(0),
\end{align} 
where a state observable $\caO$ was considered.  
Finally, using the explicit form of the entropy variation \eqref{deStot_sim}, we find for the response function ($t' < t$)
\begin{align}
\label{many_dof}
R_{\caO,\theta}(t,t')  = 
\sum_i \frac{\epsilon_i}{4  T_i^2} \bigg[ & \frac 1 {2 \mu_i} \frac{\diff^2}{\diff t'^2}\mean{\caO(t)x_i^2(t')} 
 -  \mean{ \caO (t) x_i(t')  \genL^{(T_i)} F_i(t') } \nonumber\\
&+ \mean{\caO(t)  \mu_i F_i^2(t')  }
- 2 \mean{\caO(t)  \mu_i  \dot x_i(t') F_i(t')   }
 \nonumber\\ &
 +  \sum_{j=1}^N \mean{ \caO (t) [x_i \dot x_j \partial_j F_i](t') }\left(\frac{T_i}{T_j}-1\right)\bigg].
\end{align}
This equation simplifies if the system is \emph{isothermal} before the perturbation is applied, i.e., the heat reservoirs are all at the same temperature $T_j= T \, \forall j$. In this case $\frac{\delta \caS }{\delta x_i}$ vanishes and~\eqref{noifU} boils down to
\begin{align}
 \mean{ \caB \frac{\delta \caA}{\delta x_i(t')} }   =&   \frac{1}{2 \mu_i T_i}  \mean{ \caB \Big[ \ddot x_i(t')- \mu_i \genL F_i(t') \Big]  },
\end{align}
once we recognise $ \genL^{(T_i)}|_{T_j=T} = \sum_{j=1}^N \genL_j $ as the total generator of the dynamics in the complete state space. 
Consequently, for isothermal systems the response formula takes the simpler form ($t' < t$)
\begin{align}
R_{\caO,\theta}(t,t') = \frac{1}{4 T^2}\sum_i \epsilon_i  \bigg[
  \frac 1 {2 \mu_i} \frac{\diff^2}{\diff t'^2}\mean{\caO(t) x_i^2(t') } 
- \mean{ \caO (t) x_i(t')  \genL F_i(t') } & \nonumber\\
 +  \mean{\caO(t) \mu_i F_i^2(t') }
  -2  \mean{\caO(t)  \dot x_i(t') F_i(t')} &
 \bigg] ,\label{R3}
\end{align}
which is a straightforward generalization of \eqref{R1} to a many-body system.

As noted above, if $\caO$ is a path-dependent observable one needs to include in the response formula the additional term 
\begin{align}
- 2 \mu_i T_i \mean{\frac{\delta \caO(t)}{ \delta x_i(t')} x_i(t')},
\end{align}
 coming from the first summand of \eqref{identity}. For the example of the total heat flux into the reservoirs, ${\caQ[\vec x]\equiv\sum_{j=1}^N \int_0^t  \diff s F_j(s) \dot x_j(s)}$, the supplementary term contains 
\begin{align}
\frac{\delta \caQ(t)}{ \delta x_i(t')}&=\sum_{j=1}^N \big(\partial_i F_j(t') - \partial_j F_i(t')\big)\dot x_j(t')\,,
\end{align}
and thus vanishes when the interactions derive from a two-body potential.

\subsection{Susceptibility}
Upon integration of \eqref{many_dof} we get an equation for the susceptibility of the system,
\begin{align}
\label{chi}
\chi_{\caO,\theta}(t) \equiv & \int_0^t \diff t' R_{\caO,\theta}(t,t') = S_1 + S_2 + K_1 + K_2
\end{align}
with
\begin{subequations}
\begin{align}
\label{chi_S1}
S_1 &= -  \mean{  \caO (t)\, 
\sum_i \frac{\epsilon_i}{2 T_i^2}\int_0^t d t' \dot x_i(t') F_i(t') } 
\\
\label{chi_S2}
S_2 &= 
\mean{ \caO (t) \, 
\sum_i \frac{\epsilon_i}{4 T_i^2} 
\sum_{j=1}^N \left(\frac{T_i}{T_j}-1 \right) \int_0^t d t'  [x_i \dot x_j \partial_j F_i](t') } 
\\
\label{chi_K1}
K_1 &=  \mean{\caO(t)\,  
\sum_i \frac{\epsilon_i}{4 T_i^2} \int_0^t d t' \left[ 
 \mu_i F_i^2(t')  + x_i(t')  \genL^{(T_i)} F_i(t') \right]}
\\
\label{chi_K2}
K_2 & = \left.\frac{\diff}{\diff t'}\mean{\caO(t) \,
\sum_i \frac{\epsilon_i}{8 \mu_i T_i^2} x_i^2(t')}\right|_{t'=0}^{t'=t}
\end{align}
where we recall that integrals are in the Stratonovich sense and $\genL^{(T_i)}$ was introduced in \eqref{genLT}.
The term $S_1$ is the standard correlation between observable and entropy production, appearing with a $1/2$ prefactor with respect to the equilibrium version (see next section), in which it would be the only correlation relevant for determining the linear response. The term $S_2$ is a novel correlation between observable and a time-antisymmetric quantity, proportional to the functional variation of the bath entropy $\frac{\delta \caS[\vec x]}{\delta x}$, which may be non-zero only if $T_j \ne T_i$ for some $j$.
The remaining correlations, the {\em frenetic} terms~\cite{bai13} $K_1$ and $K_2$, collect correlations between the observable and time-symmetric dynamical features.
As in previous studies of force perturbations, both $S$'s and $K$'s contain, respectively, the entropy and {\em frenesy}~\cite{bai13} in excess due to the perturbation. 

In order to correctly evaluate the time derivative of the correlation in $K_2$, when dealing with data it is important to avoid taking discrete-time derivatives with $t'>t$ because cusps are not unusual in correlation functions for $t'\to t$. To compute numerically $\frac{d}{d t'}\mean{x_i^2(t') \caO(t)}|_{t'=t}$, in the examples of the following section we have estimated the slope of data for $\mean{x_i^2(t') \caO(t)}$ with $t' \lesssim t$.

Only if averages are evaluated in a steady state, $K_2$ can be modified as
\begin{align}
\label{chi_K2ss}
K_2^{s} & =   \mean{ \genL \caO(t) \sum_i \frac{\epsilon_i}{8 \mu_i T_i^2}[x_i^2(0)-x_i^2(t)]} 
\end{align}
\end{subequations}
because
\begin{align}
\frac{d}{d t'}\mean{\caO(t) x_i^2(t') }\bigg|_{t'=0}^{t'=t}
= -\frac{d}{d t}\mean{\caO(t) x_i^2(t') }\bigg|_{t'=0}^{t'=t}
= -\mean{\genL \caO(t) x_i^2(t')}\bigg|_{t'=0}^{t'=t}
\end{align}
A steady state susceptibility $\chi^s_{\caO,\theta}(t)  = S_1 + S_2 + K_1 + K_2^s$ is associated with $K_2^s$.

\subsection{A steady state formula and its reduction to the Kubo formula at equilibrium}\label{subsec:Kubo}
Every thermal response formulation should reduce to the standard Kubo formula
when the system is under complete equilibrium conditions at temperature $T$. 
These conditions are met if conservative forces $F_i=- \partial_i U$ (with $U(\vec x)$ the system's energy) are present, if $T_i=T \,\, \forall i$ and the perturbation is applied to a thermalised system, namely $\rho_0(\vec x)$ is the canonical distribution at temperature $T$. 
In equilibrium, the Kubo formula expresses the response function as
\begin{align}\label{FDT}
R^{\rm Kubo}_{\caO, \theta}(t-t')  =& \frac{1}{T^2} \frac{d}{d t'}\mean{\caO(t) U(t')},
\end{align}
and the corresponding susceptibility is
\begin{align}\label{FDT_chi}
\chi^{\rm Kubo}_{\caO, \theta}(t)  =& \frac{1}{T^2} \mean{\caO(t) [U(t)-U(0)]},
\nonumber\\
 =& \frac{1}{T^2} \mean{\caO(t) \caQ(t)},
\end{align}
where $\caQ(t) = U(t)-U(0)$ is the heat transferred to the system in the time interval $[0,t]$.
This formula shows that the temperature response in equilibrium is totally determined by the correlation between observable and the entropy $\caQ(t)/T$ {\em paid} by the reservoir to change the system energy.

When a global perturbation is applied to an isothermal steady state regime, say with $\epsilon_i=1 \,\, \forall i$, eq.~\eqref{R3} may be recast in an alternative form, that correctly reduces to the Kubo formula~\eqref{FDT} in equilibrium, as we show in the following. In the derivation we stay in a generic steady state condition until the very end, so that in turn we obtain another quite general formula for the response function, eq.~\eqref{R_with_v} below, in which the genuine nonequilibrium contribution is well distinguished from the Kubo correlation. A possible practical issue of such elegant separation is that it can be computed explicitly only if one knows the microscopic probability density of states.

We start noticing that the last term in~\eqref{R3} is in equilibrium half of the expected result:
\begin{align}
-\frac{1}{2 T^2} \sum_i \mean{\caO(t) \dot x_i(t')\partial_i U (t')} =\frac{1}{2 T^2}  \frac{d}{d t'} \mean{\caO(t) U(t')}. \label{half}
\end{align}
The remaining frenetic terms yield an analogous contribution at equilibrium. To show that, we first use that the system is in a stationary state. This implies that correlations are functions of the time difference only, hence $\frac{d}{d t'}$ can be exchanged with $-\frac{d}{d t}$. Moreover, the backward generator can be expressed in terms of the generator of the time-reversed dynamics, $\genL^*$, through the relation $\genL=\genL^*+ 2 \sum_{j=1}^N v_j \partial_j$, where $\vec v_j \equiv J^s_j/\rho^s $ is the state velocity, that is the probability current $J^s_j$ associated to $x_j$, over the steady state density of the system $\rho^s$ \cite{che09,bai13}.
We will ultimately exploit the time-reversal invariance of equilibrium states, which formally manifests in the equality $\genL=\genL^*$, as the probability currents $v_j$ are by definition absent at equilibrium.

The time derivatives in \eqref{resp} can be manipulated as
\begin{align}
  \frac{1}{2 \mu_i}\frac{d^2}{d t'^2}\mean{x_i^2(t') \caO(t)}&= -  \frac{1}{2 \mu_i}\frac{d}{d t'} \frac{d}{d t}\mean{x_i^2(t') \caO(t)} \nonumber\\
 &=  - \frac{1}{2 \mu_i}\frac{d}{d t'}  \mean{x_i^2(t') \genL \caO(t)} \nonumber \\
 &=  - \frac{1}{2 \mu_i}\frac{d}{d t'}  \mean{x_i^2(t') \left(\genL^*+ 2\sum_j v_j \partial_j \right)\caO(t)}  \nonumber\\
 &=  - \frac{1}{2 \mu_i}\frac{d}{d t'} \left[ \mean{ \left(\genL x_i^2(t') \right) \caO(t)} + 2 \sum_j\mean{x_i^2(t') v_j \partial_j \caO(t) } \right] \nonumber \\
& = - \frac{d}{d t'} \left[\mean{x_i(t')F_i(t') \caO(t)} + 2\sum_j \mean{x_i^2(t') v_j \partial_j \caO(t) } \right] 
 \label{first_quarter}
\end{align}
Together with stationarity, we used that $\genL^*$ is the adjoint of $\genL$, and the equality $\genL x_i^2= 2\mu_i F_i x_i + \text{const}$  in the last passage. 
We then turn to the second and third summand in~\eqref{resp}, starting with the rewriting $\mu_i F_i^2=  F_i\genL x_i$:
\begin{align}
\mean{\caO(t) \left( F_i(t') \genL x_i(t') -  x_i(t')  \genL F_i(t') \right)}&=   \mean{\caO(t) \left( F_i(t') [\genL, x_i](t') -  x_i(t') [ \genL ,F_i] (t') \right)}  \nonumber \\
&= - \mean{\caO(t) \left( F(t')  \dot x(t') -  x(t') \dot F (t') \right)} . \label{second_quarter}
\end{align}
Here we introduced the commutator acting as, e.g., $[x_i,\genL] \equiv x_i\genL- \genL x_i$, and exploit the fact that in the operator formalism time derivatives within average values are given by $\dot \caO = [\caO, \genL]$, for any state observable $\caO$ (see~\ref{sec:comm}). 
Putting together equations~\eqref{half},~\eqref{first_quarter} and \eqref{second_quarter} we obtain an expression of the thermal response valid under stationary isothermal conditions,
\begin{align}
\label{R_with_v}
R_{\caO, \theta}(t,t')  =-
\frac{1}{T^2}  \sum_i \bigg[ \mean{O(t) \dot x_i(t') F_i(t')}  
+ \frac{1}{ \mu_i} \frac{d}{d t'}\mean{\sum_j v_j(t) \partial_j O(t) \, x^2_i(t') }   \bigg].
\end{align}
Finally, at equilibrium the Kubo formula \eqref{FDT} is retrieved by setting $v_j=0 \,\,\forall j$ and using the rewriting \eqref{half} for potential forces.
Equation \eqref{R_with_v} is a thermal response counterpart of previous results for the steady-state force response based on the notion of state velocity~\cite{spe06,che08}.

\section{Examples}\label{sec:num}

\subsection{Specific heat for a quenched toy system}

In this first example we want to highlight that this framework is valid not only for steady states but also for transient regimes. There is to recall an understood dependence of the statistical averages $\mean{\ldots}$ on the initial density of states $\rho_0$. 

Let us consider a paradigmatic model of nonequilibrium overdamped systems, namely a single particle in a periodic potential $U(x) = \cos x$ and subject to an additional constant force $f$, for simplicity with mobility $\mu=1$. Thus  $F(x) = \sin x + f$, in the evolution equation \eqref{ods} of the unperturbed system. The backward operator acts on the force as $\genL F(x) = \sin x \cos x - T \sin x$.

To generate a transient condition we choose to thermalise the particle at $T_0\ne T$ and to switch to $T$ only at $t=0$, when the perturbation is also applied. In this way, even for $f=0$ one cannot apply the Kubo formula for equilibrium systems, as the initial state in not in equilibrium at temperature $T$.
Due to the periodic potential, as an arbitrary procedure for obtaining a well defined $\rho_0(x)$, we shift  to the interval $[0,2 \pi]$ any $x$ obtained from a long simulation run. However, averages such as $\mean{x^2(t') \caO(t)}$ need to be computed with $x$ interpreted as a non-periodic coordinate.
We adopted a Heun scheme~\cite{sek10} to integrate the stochastic equation, because it yields trajectories that are consistent with the Stratonovich path-weights used in our theory.

\begin{figure}[!t] 
\begin{center}
\includegraphics[width=0.48\textwidth]{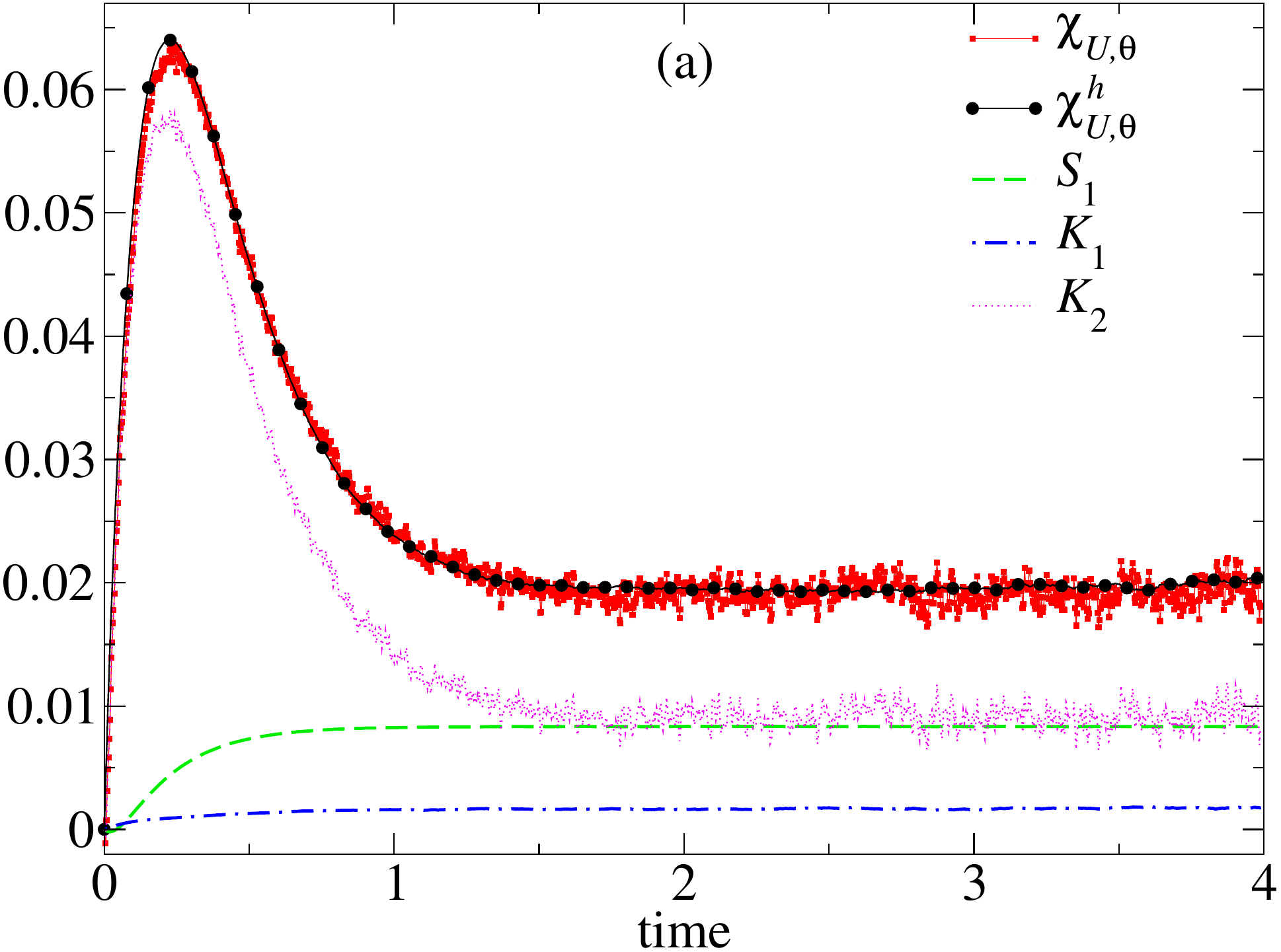}
\includegraphics[width=0.48\textwidth]{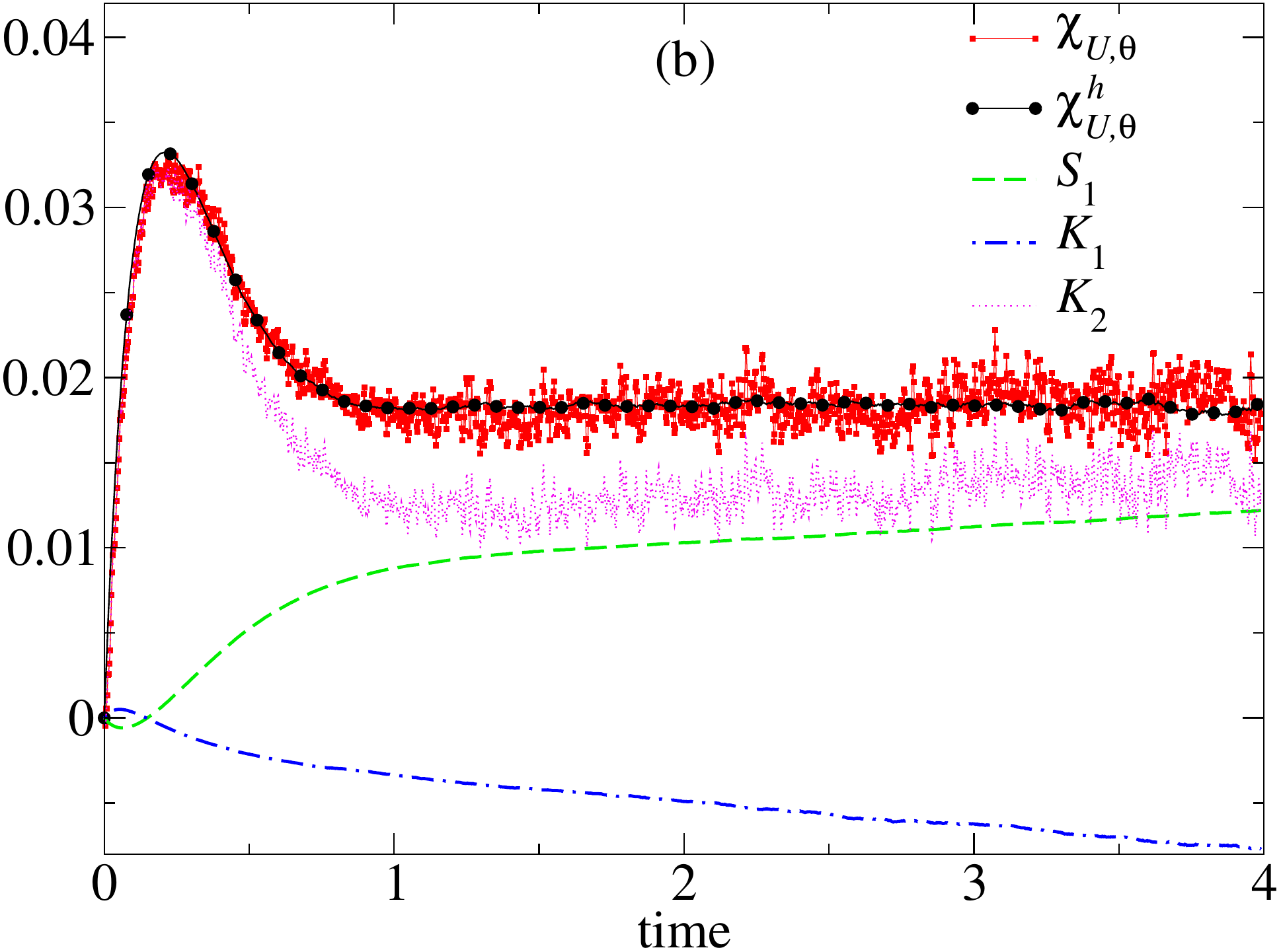}
\end{center}
\caption{Temperature susceptibility of the energy $U(x) = \cos x$ of a single particle, computed with the formula $(\chi)$ and by actually perturbing the system $(\chi^h)$. Also the single terms of the formula are shown.
The system is out of equilibrium because of a quench  at time $t=0$ from an initial $T_0=5$ to $T=0.3$. 
Consistently, the response is not given by twice the correlation $S_1$ between entropy produced and observable.
In (a) there is no additional constant force ($f=0$), while $f=0.7$ in (b) generates a nonequilibrium steady state previous to the quench.
Averages are over $4 \times 10^7$ trajectories, integrated with finite time step $dt=2.5 \times 10^{-3}$.
}
\label{fig:1p1} 
\end{figure}

In figure \ref{fig:1p1} we show examples of susceptibilities of the internal energy ($\caO = U$) to a change of $T$ for $T_0=5$ and $T=0.3$, both for $f=0$ and $f=0.7$. We compare the susceptibility $\chi_{U,\theta}(t)$ from \eqref{chi} with that computed directly as
\begin{equation}
\chi_{U,\theta}^h(t) = \frac{\mean{U(t)}_{\theta=h} - \mean{U(t)}_{\theta=0}}{h}
\end{equation}
with $h = T/100$ active from $t=0$ on. We note that, for $f=0$, the force $F$ is potential and thus the heat exchanged with the bath reduces to an energy difference,  $\caQ= -\int_0^t d t' \partial_x U(t') \dot x(t')= U(0)-U(t)$. Therefore, the susceptibility of the energy gives in the long-time limit the specific heat $C$ of the system:
\begin{align}
C\equiv  -\lim_{t\to \infty} \int_0^t d t' \frac{\delta \mean{\caQ(t)}_{\theta}}{\delta \theta(t')}
\Bigg|_{\theta=0}=
 \lim_{t\to \infty} \chi_{U,\theta}(t) 
\end{align}
If a Kubo formula \eqref{FDT} were valid, twice the entropic term \eqref{chi_S1} would yield the response. One can note that this is not the case, rather all terms in the response formula are relevant for determining the correct form of the susceptibility. In these examples, in particular, the term \eqref{chi_K2} is especially important. Being the derivative of a correlation function, it is however the noisiest one. One could resort to some high-frequency filtering for better results. In the example of the following subsection we will show that \eqref{chi_K2ss} is a good alternative to \eqref{chi_K2} in case one is dealing with steady states.

\subsection{Thermal expansion in a temperature gradient}

In equilibrium at a given temperature $T$, the correlation function between the heat absorbed by a system and its
length may be used to predict the thermal expansion response. In this example we show how this picture breaks down
out of equilibrium, where, as exposed in the previous sections, one needs to know also correlations between
length and time-symmetric observables, given by \eqref{chi_K1} and \eqref{chi_K2} or \eqref{chi_K2ss}, as well as the new entropic form \eqref{chi_S2} due to temperature unbalances. This example specialises to steady state conditions but, with respect to the previous examples, it includes the more general setup of multiple heat baths, in which one can exploit the general formulation with perturbation amplitudes $\epsilon_i$. 

\begin{figure}[!t] 
\begin{center}
\includegraphics[width=0.48\textwidth]{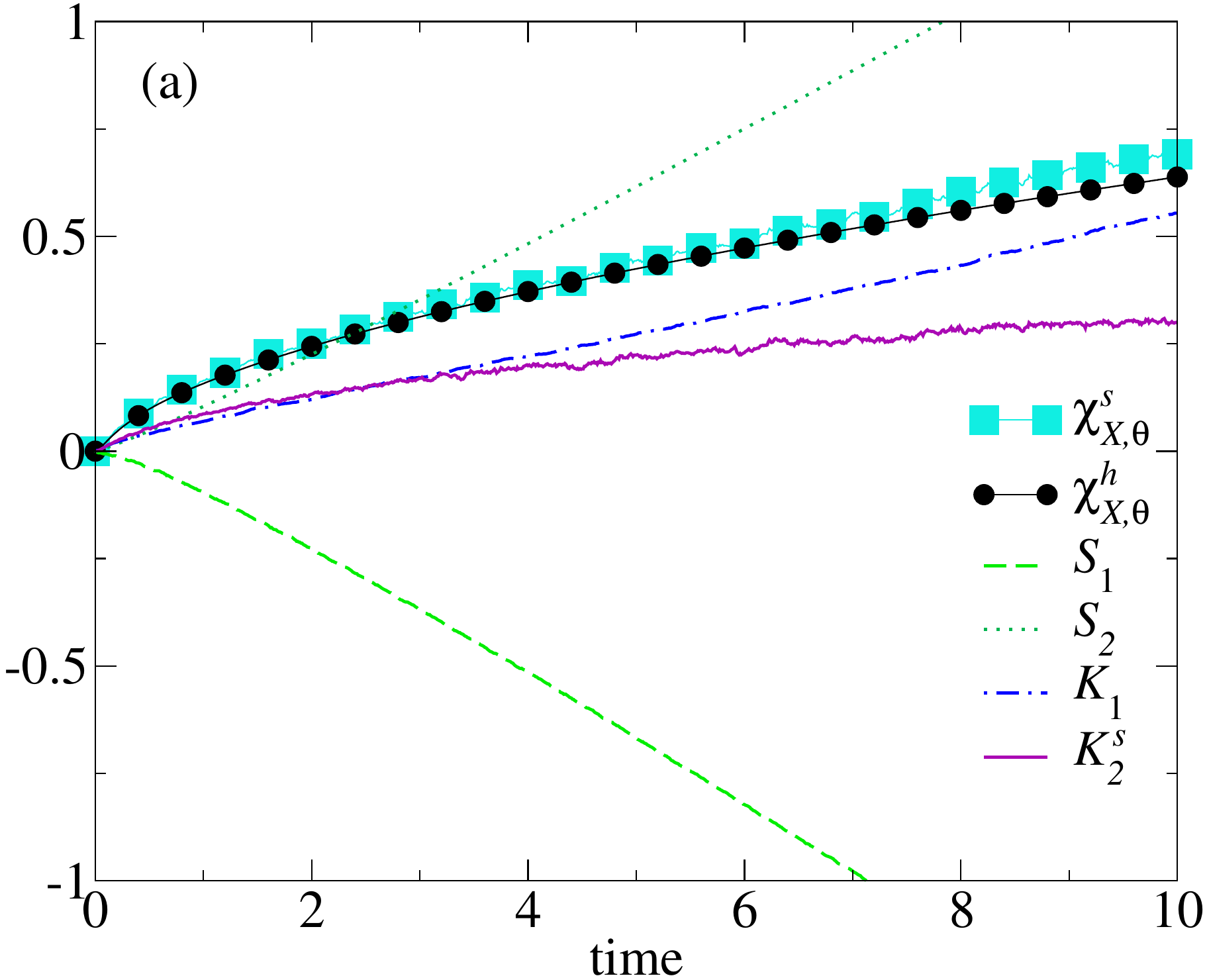}
\includegraphics[width=0.48\textwidth]{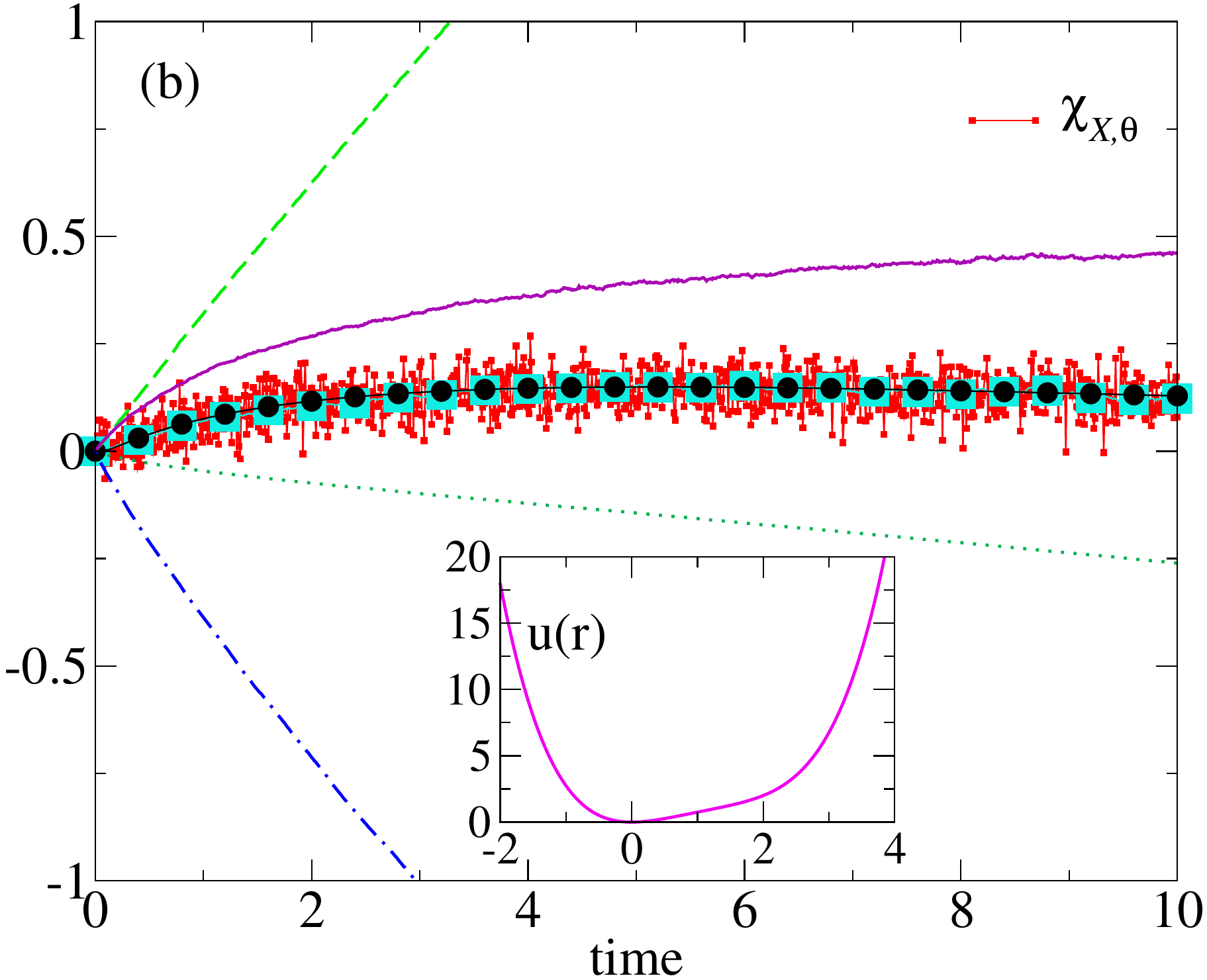}
\end{center}
\caption{Temperature steady-state susceptibility of the length $X$ of the overdamped chain ($N=11$), computed with the formula $(\chi^s)$ and by actually perturbing the system $(\chi^h)$. Also here the single terms of the formula are displayed. In these examples, $T_i$ varies linearly from $T_1=1$ to $T_N=2$.
In (a) the response is to a global temperature rise, while in (b) it is to an increase of the gradient $T_N-T_1$ preserving the average bath temperature (the inset shows the interaction potential).
Averages are over $10^7$ trajectories, integrated with finite time step $dt=10^{-3}$.
}
\label{fig:chain} 
\end{figure}

Let us consider the $N$ degrees of freedom arranged in a one-dimensional chain. The system has an energy
\begin{equation}
U(\vec x) = \frac {x_1^2}2 + \sum_{i=1}^{N-1} u(x_{i+1}- x_i), 
\qquad\textrm{with}\quad u(r) = \frac{(r-1)^4}4 + r -\frac 1 4
\end{equation}
which determines the forces, $F_i(\vec x)= -\partial_i U(\vec x)$, and again mobilities $\mu_i$ are set equal to $1$ for simplicity. 
The $x_1^2/2$ term is a pinning potential on the first site, and $x_i$'s represent the displacements from
the average positions. The length of the system in excess with respect to the length at zero temperature, $X\equiv x_N-x_1$, increases on average for increasing $T_i$'s due to the asymmetric two-body potential $u(r)$ (see the inset of figure~\ref{fig:chain}(b)).
As a paradigm of nonequilibrium conditions, the system is driven by a set of temperatures varying linearly from
$T_1$ to $T_N>T_1$.

We study the response of the length $X$ to temperature variations, in the form of (a) a global constant increase of the temperatures given by a constant $\epsilon_i=1$, and (b) an increment of the gradient $T_N-T_1$, chosen so that the average temperature is unaltered by varying $\epsilon_i$ linearly from $\epsilon_1=-1$ to $\epsilon_N=1$. For both cases, in figure~\ref{fig:chain} we see that the susceptibility $\chi^s_{X,\theta}$ computed with the steady state term \eqref{chi_K2ss} agrees fairly well with the direct estimate of the response,
\begin{equation}
\chi_{X,\theta}^h(t) = \frac{\mean{X(t)}_{\theta=h} - \mean{X(t)}_{\theta=0}}{h},
\end{equation}
 obtained with a constant $h=0.005$ turned on at $t=0$.
From figure~\ref{fig:chain} one also sees that the entropic and frenetic terms have opposite trends, between each other and with switched roles in the two cases, complementing each other to sum up to the correct response level.
In figure~\ref{fig:chain}(b) we also show the response  $\chi_{X,\theta}$ obtained by an evaluation of the time-derivative in \eqref{chi_K2} (the local variation in time of the correlation function is obtained through a linear fit of data relative to four nearby time steps). It results more noisy than the estimate via $\chi^s_{X,\theta}$.

\subsection{Free diffusion of one degree of freedom}\label{sec:free}

Let us consider the equations of motion \eqref{ods} for free diffusion of a single degree of freedom,
 $\dot x(t)=\hat \xi(t)$ with $\hat \xi=\sqrt{2 \mu T} \xi$.
The noise prefactor $\sqrt{2 \mu T}$ comes from assuming the bath to be in equilibrium.
In this way the mean square displacement of a free particle in a time $t$ is simply 
$\mean{x^2(t)} = 2 \mu T t \equiv 2 D t$, the response of the mean velocity to a small force is the free-particle
mobility $\mu$, and the Einstein relation $\mu = D/ T$ between diffusion constant $D$ and mobility is found.
One can note that the susceptibility of the observable $\caO(t) = x^2(t)$ to a change of $T$ is expected to be $2 \mu t$, hence the corresponding response function is $2\mu$. We show how our formalism reduces to this result.

For free diffusion all terms in~\eqref{R1} drop but the one involving the second derivative. In this case, the response function can be calculated directly from its definition \eqref{resp} and one can thus prove analytically that both sides of \eqref{R1} are equal to the same quantity. As we argued above, the response of the mean square displacement to the perturbation $T\to\Theta(t) = T+\theta(t)$ is
\begin{align}
\frac{\delta \mean{x^2(t)}_{ h}}{\delta \theta(t')} &=\frac{\delta }{\delta \theta(t')} \Big<x_0^2  +2 x_0 \int_0^t \diff s \hat \xi(s) + \int_0^t \diff s \int_0^t \diff u \hat  \xi(s) \hat \xi(u)\Big>_{ \theta} 
\nonumber\\
& = 2\mu \frac{\delta }{\delta \theta(t')} \int_0^t \diff s \int_0^t \diff u \delta(s-u)  \Theta(s) \nonumber\\
& = 2 \mu.
\end{align}
where we used that the initial condition is independent of the perturbation and noise, thus only the noise autocorrelation contributes. On the other hand, the response formula~\eqref{R1} becomes 
\begin{align}
R_{x^2  T}(t,t') & = \frac{1}{8 \mu T^2} \frac{\diff^2}{\diff t'^2}\mean{x^2(t') x^2(t)}\\
&= \frac{1}{8 \mu T^2}\frac{\diff^2}{\diff t'^2}\left( \mean{x^2(t')}\mean{x^2(t)} +2 \mean{x(t') x(t)}^2\right),\nonumber
\end{align}
making use of Wick's theorem to split the 4-point correlation into products of 2-point correlations. The latter read
\begin{align}
\mean{x(t')x(t)} &= \mean{x_0^2} + \int_0^{t'} \diff s \int_0^t \diff u \mean{\xi(s)\xi(u)}=  \mean{x_0^2} + 2 \mu T \min(t',t),
\end{align}
leading to a result in agreement with the previous calculation:
\begin{align}
 R_{x^2  T}(t,t') &=  \frac{1}{8 \mu T^2} \frac{\diff^2}{\diff t'^2}\Big[ 
\left(  \mean{x_0^2} + 2 \mu Tt'  \right)  \left(  \mean{x_0^2} + 2 \mu Tt  \right)+ 2 \left(  \mean{x_0^2} + 2 \mu Tt'  \right)^2 \Big] \nonumber \\
&= \frac{1}{8 \mu T^2} \frac{\diff^2}{\diff t'^2}\Big[  3\mean{x_0^2}^2 + 2\mu T  \mean{x_0^2}( t' + t ) + (2\mu T)^2 t t' + 8 \mu T t'  \mean{x_0^2} + 2 (2 \mu T)^2 {t'}^2 \Big] \nonumber\\
&= 2 \mu.
\end{align}
As expected, interchanging $\frac{\diff }{\diff t'^2}$ with  $\frac{\diff }{\diff t^2}$ would give an incorrect result as the system is not in a steady state. It is also trivial to verify~\eqref{R2}, namely that the this result coincides with the second order response to a state-independent force, giving rise to the dynamics $ \dot x(t)=\mu f(t)+\hat \xi(t)$. Indeed, using again the conditions of independency of the initial condition, one finds
\begin{align}
\frac 1 \mu \frac{\delta^2 \mean{x^2(t)}_{ f}}{\delta  f^2(t')} 
& = \frac 1 \mu \frac{\delta^2 }{\delta f^2(t')}  \mean{\left(\int_0^t \diff s \big(\mu f(s)- \hat \xi(s)\big) \right)^2}\nonumber\\
& = \mu \frac{\delta^2 }{\delta  f^2(t')} \int_0^t \diff s \int_0^t \diff u f(s) f(u) 
\nonumber\\&
 = 2 \mu.
\end{align}

\section{Conclusions}
For overdamped stochastic systems far from equilibrium we have obtained the linear response function of generic state observables to a change in the temperature of the Langevin heat baths. Improving a previous result \cite{bai14}, we need not express the response in terms of a finite time mesh, being all the divergencies appearing in the continuous limit removed, and being all terms in the susceptibility standard integrals or derivatives. This was achieved by deriving a sort of Dyson-Schwinger equation \cite{zinn02}, i.e., a relation between unperturbed correlation functions involving an arbitrary observable. This method complements and expands our recent results \cite{fal15b} obtained via a different approach, in which the additional noise stemming from the perturbation was turned into mechanical forces by means of a space rescaling.

As in many previous examples, in order to describe a nonequilibrium systems, one needs to know more than just the
entropy production.
The additional information concerns the knowledge of dynamical quantities that are even under the reversal of the arrow of time (squares of forces, etc.). Among them we have recognized the change of the time-integral of the effective potential (i.e., the total escape rate integrated along trajectories) upon variation of the perturbed degree of freedom, $\frac{\delta \caK}{ \delta x_i}$. This quantity emerges from the regularization procedure we set up, along with the change of the total bath entropy flow $\frac{\delta \caS}{ \delta x_i}$, which complements, perhaps surprisingly, the usual entropy production entering Kubo formula.

For the common scenario of isothermal systems in a steady state, we have also shown how to convert the results in a formula that separates the Kubo term from a nonequilibrium additional correlation that includes the state velocity, see~\eqref{R_with_v}. Such version is complementary to the others in the sense that it requires the knowledge of the density of states rather than that of dynamical details.

Future developments of this framework should include multiplicative noise, i.e. those cases where the temperature experienced by the particle depends on their positions.

\appendix

\section{Derivation of the second order response function}\label{sec:HS}
The derivation of~\eqref{R2} starts with a Hubbard-Stratonovich transformation of the path weight, which is a functional generalization of the integral identity $\int \diff y e^{-Dy^2-{\rm i} z y}= e^{-\frac{z^2}{4D}} \sqrt{\frac \pi D}$ 
(below the $\sqrt{\pi/D}$ is adsorbed in the path measure $\caD y$)
valid for real $y$ and $D>0$. When applied to~\eqref{path} and~\eqref{action} it renders the response \eqref{resp} in the form \eqref{R2} through the following manipulations:
\begin{align}
 \frac{\delta\mean{ \caO(t)}_{\theta}}{\delta \theta(t')}\Bigg|_{\theta=0} = 
& \frac{\delta}{\delta \theta(t')} \int \caD \vec x \diff \vec x_0 \caD \vec y \rho_0(\vec x_0) \caO(t) \times \nonumber\\
& \quad \prod_{j=1}^N \exp \bigg\{ - \int_{0}^{t} \diff s \bigg[ 
  \mu_j \Theta_j y_j^2 -  \mathrm{i} y_j (\dot x_j - \mu_j F_j)
  + \frac{1}{2}\mu_j \partial_j F_j \bigg]\bigg\} \Bigg|_{\theta=0} 
\nonumber\\
= 
& \int \caD \vec x \diff \vec x_0 \caD \vec y \rho_0(\vec x_0) \caO(t) \bigg[-\sum_i\epsilon_i \mu_i y_i^2(t')\bigg]
\times
\nonumber\\ 
& \quad \prod_{j=1}^N\exp\bigg\{
  - \int_{0}^{t} \diff s \bigg[
  \mu_j T_j y_j^2 -  \mathrm{i} y_j (\dot x_j - \mu_j F_j)  +\frac{1}{2}\mu_j \partial_jF_j\bigg]\bigg\}  \label{y^2}\\
=
& \sum_i\frac{\epsilon_i}{\mu_i}\frac{\delta^2}{\delta f_i ^2(t')}
  \int \caD \vec x \diff \vec x_0 \caD \vec y \rho_0(\vec x_0) \caO(t) 
\times 
\label{forceh}\\
& \quad \prod_{j=1}^N\exp\bigg\{ - \int_{0}^{t} \diff s \bigg[
  \mu_j T_j y_j^2 - \mathrm{i}y_j (\dot x_j - \mu_j F_j-\mu_j f_j)
  + \frac{1}{2}\mu_j \partial_j F_j \bigg]\bigg\}\Bigg|_{\vec f=0} \nonumber\\
=
&
\sum_i\frac{\epsilon_i}{\mu_i}R^{(2)}_{\caO,f_i}(t,t',t'),  
\nonumber
\end{align}
where we rewrote \eqref{y^2} introducing the derivatives of a state-independent force $f_i$, and recognised in \eqref{forceh} the Martin-Siggia-Rose path-weight \cite{mar73} associated to the perturbed dynamics \eqref{const}.

\section{Variation of the action functional}\label{sec:path_rel}
Here we detail the calculation of the functional variation of the path-weight action $\caA[\vec x]$ that was used in Section~\ref{sec:reg}. For the sake of clarity we distinguish the single-particle from the many-particle case. 
\subsection{One degree of freedom}
For $N=1$, the action is given by \eqref{S} and its variation is
\begin{align}\label{deA}
 \frac{\delta \caA}{\delta x(t')}  &= \frac 1 2   \frac{\delta  \caS  }{\delta x(t')}
- \frac{\delta  \caK  }{\delta x(t')}  +  \frac{\ddot x (t')}{2 \mu T}\,.
\end{align}
The variation of the bath entropy is identically zero, unless $F$ is an explicit function of time $F(t')=F(x(t'),t')$:
\begin{align}
\frac{\delta \caS}{ \delta x(t')}&=\frac{1}{T}\left(\partial_x F(t') \dot x(t')+ \int_0^t \diff s \dot \delta(s-t') F(s)\right) \nonumber\\
&=\frac 1 T \left(\partial_x F(t') \dot x(t')-\partial_{t'} F(t')- \partial_x F(t') \dot x(t') \right)=- \frac 1 T \partial_{t'} F(t')\,. \nonumber
\end{align}
Since the  dynamical activity is independent of $\dot x$, its variation is simply the derivative of the escape rate from $x(t')$:
\begin{align}\label{deK}
\frac{\delta \caK}{ \delta x(t')}&= 
\partial_x \lambda(t')=\frac 1 {2 T} \left(\mu F(t')\partial_x F(t') - \mu T \partial_x^2 F(t') \right).
\end{align}
Therefore, introducing in \eqref{deK} the backward generator $\genL$, \eqref{deA} becomes
\begin{align}
  \frac{\delta \caA}{\delta x(t')} 
&= \frac{1}{2\mu T}   \Big[ \ddot x(t')- \mu \genL F(t')   - \mu \partial_{t'}F(t')\Big].\label{derivative} 
\end{align}
As a side note, plugging this result into \eqref{identity} with $\caB=1$ one obtains (if $F$ deepens on $x$ only)
\begin{align}
\mean{\ddot x}= \mu\mean{\genL F},
\end{align}
i.e., the mean trajectory satisfies Newton's equation with an effective force $\mu \genL F$. In the weak-noise limit $T\ll 1$, such trajectory becomes the most probable one, being the minimiser of the action.
This expression could be obtained directly by applying the backward generator $\genL$ to the Langevin equation \eqref{ods}, and using that $\xi$ does not depend on $x$.
\subsection{Many degrees of freedom}
For $N>1$, thanks to the independency of the different thermal noises, the action \eqref{action_un} is simply the sum of ``single-coordinate'' actions: $\caA[\vec x]= \sum_{j=1}^N \caA^{(j)}[\vec x]$ with  $\caA^{(j)}$ following the structure \eqref{S}. Nevertheless, its variation is not just equal to \eqref{derivative} but in general it will contain additional terms owing to the interactions between different degrees of freedom. One indeed finds modified expressions for the variation of the total entropy flux into the (unperturbed) reservoirs,
\begin{align}
\frac{\delta \caS[\vec x] }{\delta x_i(t')} &= \sum_{j=1}^N  \dot x_j(t')\left(\frac{\partial_i F_j(t')}{T_j} -\frac{\partial_j F_i(t')}{T_i} \right) - \frac{1}{T_i}\partial_{t'}F_i(t')\, ,
\end{align}
and for the variation of the total dynamical activity
\begin{align}\label{deK_b}
\frac{\delta \caK}{ \delta x(t')}&= \sum_{j=1}^N\partial_i \lambda_j(t')= 
\sum_{j=1}^N\frac{1}{ 2 T_j } \left( \mu_j F_j(t')\partial_i F_j(t') +  \mu_j T_j \partial_i\partial_j F_j(t') \right),
\end{align}
which in general  cannot be cast in terms of the total backward generator $\genL$.
The variation of the action is thus given by
\begin{align}
 \frac{\delta \caA}{\delta x_i(t')}  
& =  \frac 1 2   \frac{\delta  \caS  }{\delta x_i(t')}
- \frac{\delta  \caK  }{\delta x_i(t')}  +  \frac{\ddot x_i (t')}{2 \mu_i T_i} \nonumber\\
& =   \frac{1}{2 \mu_i T_i} \bigg[ \ddot x_i(t')- \mu_i T_i \sum_{j=1}^N\partial_i \lambda_j(t')- \mu_i \partial_{t'}F_i(t') \nonumber \\
& \qquad \qquad +  \mu_i T_i   \sum_{j=1}^N  \dot x_j(t')\left(\frac{\partial_i F_j(t')}{T_j} -\frac{\partial_j F_i(t')}{T_i} \right) 
 \bigg] .  \label{deStot} 
\end{align}
Equation \eqref{deStot} is completely general, and thus, when combined with \eqref{noifU},  provides a regularised expression for the thermal response of overdamped systems under any nonequilibrium conditions:
\begin{align}
\label{general}
R_{\caO,\theta}(t,t')  = \sum_i \frac{\epsilon_i}{4  T_i^2} \Bigg[
& \frac{1}{\mu_i}
\frac{\diff^2}{\diff t'^2}\mean{\caO(t)x_i^2(t')} - \mean{ \caO(t) x_i(t')\partial_{t'}F_i(t') }\nonumber\\
& +  \mean{ \caO (t) \left(  \mu_i F_i^2(t')-2  \dot x_i(t') F_i(t')   -x_i(t')  T_i
\sum_{j=1}^N \partial_i \lambda_j(t')  \right)} \nonumber\\
&+  T_i  \sum_{j=1}^N \mean{ \caO (t) x_i(t') \dot x_j(t') 
\left(\frac{\partial_i F_j(t')}{T_j} -\frac{\partial_j F_i(t')}{T_i} \right)}\Bigg].
\end{align}

Nevertheless, the cross-terms $\partial_i F_j$ with $i \neq j$ appearing in~\eqref{deStot} simplify considerably if we assume that the degrees of freedom interact with each others via a two-body potential $\caU(\{x_i-x_j\})$. Hence we can exploit the relation
\begin{align}\label{simply}
\partial_i F_j= - \partial_i \partial_j \caU= - \partial_j \partial_i \caU=\partial_j F_i,
\end{align}
which is nothing but the action-reaction principle. Equation \eqref{deStot} then becomes 
\begin{align}
\frac{\delta \caA}{\delta x_i(t')}  &= \frac{1}{2 \mu_i T_i}\bigg[ \ddot x_i(t')- \mu_i\genL^{(T_i)} F_i(t')- \mu_i \partial_{t'}F_i(t') \bigg] +  \sum_{j=1}^N   \dot x_j(t') \partial_j F_i(t')\left(\frac{1}{2T_j} -\frac{1}{2T_i} \right) \,. \label{deStot_sim}
\end{align} We remark that for systems in $d=1$ \eqref{simply} does not impose any limitation on the driving, that is, one-body non-conservative forces can be present as well, they simply do not enter in~\eqref{deStot}, which concerns only the interactions between different particles. Instead, in $d>1$, different indexes $i$ and $j$ in \eqref{deStot} may refer to the coordinates of the same particle, thus \eqref{deStot} cannot be simplified to \eqref{deStot_sim} in the presence of generic non-conservative forces.  

It is worth noting that when the equality $\partial_j F_i=\partial_i F_j$ holds, the choice $\caB=1$ in the identity \eqref{identity} yields the effective Newton's equation for the mean trajectory
\begin{align}
\mean{\ddot x_i}=\mu_i \mean{\genL^{(T)} F_i}-\mu_i T_i \mean{\frac{\delta \caS }{\delta x_i}}.
\end{align}
On the other hand, direct application of the operator $\genL$ to the Langevin equation \eqref{ods} gives $\mean{\ddot x_i}=\mu_i \mean{\genL F_i}$. By comparison, one concludes that there exists a natural splitting of the effective force, namely
\begin{align}
 \mean{\genL F_i}=\mu_i \mean{\genL^{(T)} F_i}-\mu_i T_i \mean{\frac{\delta \caS }{\delta x_i}},
\end{align}
where the first component originates from variations of the force $F_i$ in thermal time, while the second is a gradient-like force in which the entropy flux into the bath acts a free-energy.

\section{Time derivative in operator formalism}\label{sec:comm}
Consider the state observables $\caO_\alpha$, that are arbitrary functions of $\vec x$. In the operator formalism, their (steady-state) evolution over a time-span $t-t'$ is given by the action of the operator $e^{\genL (t-t')}$. Therefore, the typical correlation functions we are interested in are expressed by (with $t>t'$)
\begin{align}\label{eg.corr}
\mean{\caO_3(t) \caO_2(t') \caO_1(t')}= \int d \vec x_0 \rho_0(\vec x_0) e^{\genL t'}  \caO_1  \caO_2 e^{\genL (t-t')} \caO_3,
\end{align}
where the dependence of $\caO_\alpha$ on $\vec x_0$ is omitted for brevity \cite{bai13}. In analogy to the Heisenberg picture in quantum mechanics, one may include the dependency on time in the observables by the definition $\caO_\alpha(t')\equiv e^{\genL t'} \caO_\alpha  e^{-\genL t'} $. Hence, a time derivative applied to one of the operators in \eqref{eg.corr} gives, e.g., 
\begin{align}\label{eg.corr2}
\mean{\caO_3(t) \dot \caO_2(t') \caO_1(t')}&= \mean{\caO_3(t)(\genL  e^{\genL t'}  \caO_2  e^{-\genL t'}-  e^{\genL t'}  \caO_2  e^{-\genL t'} \genL  )\caO_1(t')}\nonumber \\
&=\mean{\caO_3(t)(\genL \caO_2(t')  - \caO_2(t')   \genL )\caO_1(t')}\nonumber\\
&=\mean{\caO_3(t)[\genL ,\caO_2(t')] \caO_1(t')}.
\end{align}

\section*{References}
\bibliography{bib_noneq_150723}

\providecommand{\newblock}{}
\begin{thebibliography}{10}
\expandafter\ifx\csname url\endcsname\relax
  \def\url#1{{\tt #1}}\fi
\expandafter\ifx\csname urlprefix\endcsname\relax\def\urlprefix{URL }\fi
\providecommand{\eprint}[2][]{\url{#2}}

\bibitem{ein05}
Einstein A 1905 {\em Ann. Phys.\/} {\bf 4}

\bibitem{nyq28}
Nyquist H 1928 {\em Phys. Rev.\/} {\bf 32} 110

\bibitem{ons31}
Onsager L 1931 {\em Phys. Rev.\/} {\bf 37}(4) 405--426

\bibitem{ons31b}
Onsager L 1931 {\em Phys. Rev.\/} {\bf 38}(12) 2265--2279

\bibitem{cal51}
Callen H~B and Welton T~A 1951 {\em Phys. Rev.\/} {\bf 83} 34

\bibitem{gre52}
Green M~S 1952 {\em J. Chem. Phys.\/} {\bf 20} 1281--1295

\bibitem{gre54}
Green M~S 1954 {\em J. Chem. Phys.\/} {\bf 22} 398--413

\bibitem{kub57}
Kubo R 1957 {\em J. Phys. Soc. Jpn.\/} {\bf 12} 570--586

\bibitem{tod92}
Kubo R, Toda M and Hashitsume N 1992 {\em Statistical Physics: Nonequilibrium
  statistical mechanics\/} 2nd ed vol~2 (Springer)

\bibitem{ern69}
Ernst M~H, Haines L~K and Dorfman J~R 1969 {\em Rev. Mod. Phys.\/} {\bf 41}(2)
  296--316

\bibitem{ald70}
Alder B~J and Wainwright T~E 1970 {\em Phys. Rev. A\/} {\bf 1} 18

\bibitem{dor70}
Dorfman J~R and Cohen E~G~D 1970 {\em Phys. Rev. Lett.\/} {\bf 25}(18)
  1257--1260

\bibitem{zwa70}
Zwanzig R and Bixon M 1970 {\em Phys. Rev. A\/} {\bf 2}(5) 2005--2012

\bibitem{eva90}
Evans D~J and Morriss G~P 1990 {\em Statistical Mechanics of NonEquilibrium
  Liquids\/} Theoretical Chemistry Monograph Series (London: Academic Press)

\bibitem{eva93b}
Evans D~J and Sarman S 1993 {\em Phys. Rev. E\/} {\bf 48}(1) 65--70

\bibitem{ron07}
Rondoni L and Mejia-Monasterio C 2007 {\em Nonlinearity\/} {\bf 20} R1

\bibitem{rue09}
Ruelle D 2009 {\em Nonlin.\/} {\bf 22} 855--870

\bibitem{mar08}
{Marini Bettolo Marconi} U, Puglisi A, Rondoni L and Vulpiani A 2008 {\em Phys.
  Rep.\/} {\bf 461} 111--195

\bibitem{col12}
Colangeli M, Rondoni L and Vulpiani A 2012 {\em J. Stat. Mech.\/}  L04002

\bibitem{col14}
Colangeli M and Lucarini V 2014 {\em J. Stat. Mech.\/}  P01002

\bibitem{han78}
H\"{a}nggi P 1978 {\em Helv. Phys. Acta\/} {\bf 51} 202219

\bibitem{fal90}
Falcioni M, Isola S and Vulpiani A 1990 {\em Phys. Lett. A\/} {\bf 144} 341

\bibitem{cug94}
Cugliandolo L, Kurchan J and Parisi G 1994 {\em J. Phys. I\/} {\bf 4} 1641

\bibitem{rue98}
Ruelle D 1998 {\em Phys. Lett. A\/} {\bf 245} 220--224

\bibitem{nak08}
Nakamura T and Sasa S 2008 {\em Phys. Rev. E\/} {\bf 77} 021108

\bibitem{che08}
Chetrite R, Falkovich G and {Gaw\c{e}dzki} K 2008 {\em J. Stat. Mech.\/}
  P08005

\bibitem{spe06}
Speck T and Seifert U 2006 {\em Europhys. Lett.\/} {\bf 74} 391--396

\bibitem{spe09}
Speck T and Seifert U 2009 {\em Phys. Rev. E\/} {\bf 79} 040102

\bibitem{sei10}
Seifert U and Speck T 2010 {\em Europhys. Lett.\/} {\bf 89} 10007

\bibitem{pro09}
Prost J, Joanny J~F and Parrondo J~M 2009 {\em Phys. Rev. Lett.\/} {\bf 103}
  090601

\bibitem{ver11}
Verley G, Ch\'etrite R and Lacoste D 2011 {\em J. Stat. Mech.\/}  P10025

\bibitem{lip05}
Lippiello E, Corberi F and Zannetti M 2005 {\em Phys. Rev. E\/} {\bf 71} 036104

\bibitem{lip07}
Lippiello E, Corberi F and Zannetti M 2007 {\em J. Stat. Mech.\/}  P07002

\bibitem{bai09}
Baiesi M, Maes C and Wynants B 2009 {\em Phys. Rev. Lett.\/} {\bf 103} 010602

\bibitem{bai09b}
Baiesi M, Maes C and Wynants B 2009 {\em J. Stat. Phys.\/} {\bf 137} 1094--1116

\bibitem{che09b}
Chetrite R 2009 {\em Phys. Rev. E\/} {\bf 80} 051107

\bibitem{bai14}
Baiesi M, Basu U and Maes C 2014 {\em Eur. Phys. J. B\/} {\bf 87} 277

\bibitem{yol15}
Yolcu C and Baiesi M 2015 {\em arXiv:1512.04319\/}

\bibitem{fal15b}
Falasco G and Baiesi M 2015 {\em arXiv:1509.03139\/}

\bibitem{sek10}
Sekimoto K 2010 {\em Stochastic Energetics\/} ({\em Lecture Notes in Physics\/}
  vol 799) (Springer)

\bibitem{oksendal}
{\O}ksendal B 2003 {\em Stochastic differential equations\/} (Berlin: Springer)

\bibitem{zinn02}
Zinn-Justin J 2002 {\em Quantum field theory and critical phenomena\/} 4th ed
  (Oxford: Clarendon Press)

\bibitem{lec05}
Lecomte V, Appert-Rolland C and {van Wijland} F 2005 {\em Phys. Rev. Lett.\/}
  {\bf 95} 010601

\bibitem{mer05}
Merolle M, Garrahan J~P and Chandler D 2005 {\em Proc. Natl. Acad. Sci.\/} {\bf
  102} 10837--10840

\bibitem{gar09}
Garrahan J~P, Jack R~L, Lecomte V, Pitard E, {van Duijvendijk} K and {van
  Wijland} F 2009 {\em J. Phys. A: Math. Gen\/} {\bf 42} 075007

\bibitem{ful13}
Fullerton C~J and Jack R~L 2013 {\em J. Chem. Phys.\/} {\bf 138} 224506

\bibitem{aut09}
Autieri E, Faccioli P, Sega M, Pederiva F and Orland H 2009 {\em J. Chem.
  Phys.\/} {\bf 130} 064106

\bibitem{pit11}
Pitard E, Lecomte V and {van Wijland} F 2011 {\em Europhys. Lett.\/} {\bf 96}
  56002

\bibitem{bai13}
Baiesi M and Maes C 2013 {\em New J. Phys.\/} {\bf 15} 013004

\bibitem{lau07}
Lau A~W~C and Lubensky T~C 2007 {\em Phys. Rev. E\/} {\bf 76}(1) 011123

\bibitem{negele88}
Negele J~W and Orland H 1988 {\em Quantum many-particle systems\/} (New York:
  Perseus)

\bibitem{lip08}
Lippiello E, Corberi F, Sarracino A and Zannetti M 2008 {\em Phys. Rev. E\/}
  {\bf 78} 041120

\bibitem{gro98}
Grosche C and Steiner F 1998 {\em Handbook of Feynman path integrals\/} vol~1
  (Spring)

\bibitem{che09}
Chetrite R and {Gaw\c{e}dzki} K 2009 {\em J. Stat. Phys.\/} {\bf 137} 890--91

\bibitem{mar73}
Martin P~C, Siggia E~D and Rose H~A 1973 {\em Phys. Rev. A\/} {\bf 8}(1)
  423--437

\end{thebibliography}

\end{document}